\documentstyle[12pt]{article}
\begin {document}
\vspace{1.0truein}
\centerline{\bf{ An Extension of 
Multiple Cosmic String Solution: A Proposal}}
\vspace{1.0truein}
\centerline{M.Horta\c csu $^{\dagger *}$ and N. \" Ozdemir $^{\dagger} $} 
\vspace{.4truein}
\centerline{ ${\dagger}$ Physics Department, Faculty of Sciences and Letters}
\vspace{.3truein}
\centerline { ITU 80626, Maslak, Istanbul, Turkey}
\vspace{.3truein}
\centerline {${*}$ Physics Department, 
T\" UBITAK Research Institute for Basic Sciences, Turkey}
\vspace{2truein}
\noindent 
{\raggedright{\bf{Abstract}}
We extend the work done for cosmic strings on the perturbative calculation  
of vacuum polarization of a massless field in the space-time of 
multiple cosmic strings and show that for a more
general class of locally flat metrics 
the one loop calculation do not introduce
any new divergences to the VEV of the energy of a scalar particle or a
spinor particle. We explicitly perform the calculation for the configuration 
where we have one cosmic string in the presence of a dipole made out of 
cosmic strings both for the scalar and the spinor cases.}\\
\vspace{.1truein}
PACS number is 98.80 Cq
\vfill\eject
\newpage
\baselineskip=18pt

\section{Introduction}
Extensive work has been done on cosmic strings $^{/1}$ and other
topological defects.  A remarkable property of the field theoretical
calculations using cosmic strings as background is the existence of a finite
contribution to the energy of a scalar particle due to the presence 
of the cosmic string $^{/2} $.  In a beautiful paper  $^{/3}$ A.N. Aliev
showed that there is an intricate cancellation 
mechanism in the cosmic string
background which makes the perturbative expression finite.  The divergence
due to the one loop integration is cancelled by the zero coming from the
anti Fourier transforming the resulting expression.

In this note we want to point out to a simple property of certain metrics,
made out of two ``holomorphic'' structures, where  similar cancellation 
occurs. In this sense we extend the Aliev result to more general, still 
locally flat, metrics.  These metrics also have a special property.
If the euclidean Dirac operator is written in the background of such a metric,
it is the true square root of the Klein-Gordon operator,
a property which the
free Dirac operator has in
  flat space.  If we take a special form of these metrics , this
property also holds for the Dirac operator with the lorentzian
signature.  Thus our metrics generalize an important property of the
 free Dirac operator which exists in flat space metric.
Our metrics are all locally flat, with possible Dirac delta function type
singularities.

We introduce our new set of metrics in Section II and analyze their
properties.  Section III is devoted to the calculation of the vacuum
fluctuations of a massless
scalar particle in the background of a representative member of our
metrics.                In Section IV , we perform the similar
calculation for a spinor particle.  We conclude with a few remarks.

 \section{New metrics}
 Take the special form of the Dirac operator
\begin{equation}
D'= e^{-g_1(x_0-\gamma^0 \gamma^3 x_3)} 
(\gamma^0 \partial_0 + \gamma^3 \partial_3)
+e^{-g_2 (x_1-\gamma^1 \gamma^2 x_2)} 
(\gamma^1 \partial_1 +\gamma^2 \partial_2).
\end{equation} 
 
Here $\gamma^{\mu} $ are Dirac gamma matrices 
obeying the anticommutation relations
$[\gamma^{\mu},\gamma^{\nu} ]_{+} = 2g^{\mu \nu} $, and $g_1,g_2$ are
arbitrary smooth functions which are defined by their power expansions.
Define the d'Alembertian operator as
\begin{equation} 
 \Box =  {Tr D'D'\over {4}}.
\end{equation}
One sees that the d'Alembertian operator reads
\begin{equation} 
\Box = e^{-f_1(x_0,x_3)}\left({{\partial^{2}} \over {\partial x_0^2}}
-{{\partial^{2}} \over {\partial x_3^2}}\right)-
e^{-f_2(x_1,x_2)} \left({{\partial^2} \over {\partial x_1^2}}
+{{\partial^2} \over {\partial x_2^2}} \right) ,
\end{equation}
where
\begin{equation} 
 f_1(x_0,x_1)= {Tr \over {4}} \left(g_1(x_0+\gamma^0 \gamma^3 x_3) +
 g_1(x_0-\gamma^0 \gamma^3 x_3)\right)
\label{f1}
\end{equation}
\begin{equation}  
 f_2(x_1,x_2)= {Tr \over {4} } \left(g_2(x_1+\gamma^1 \gamma^2 x_2)+
 g_2(x_1- \gamma^1 \gamma^2 x_2) \right) .
\label{f2}
\end{equation} 
One can derive the metric that will give this d'Alembertian operator,
using the relation  $ \Box =\frac{1}{ \sqrt{-g}}\partial_\mu
g^{\mu \nu} \sqrt{-g} \partial_\nu $  .
This d'Alembertian can be derived from the metric
\begin{equation}  
ds^2= e^{-f_1(x_0,x_3)} (dx_0^2 -dx_3^2) -e^{-f_2(x_1,x_2)} (dx_1^2+dx_2^2)
\label{met}
\end{equation} 
where $f_1,f_2$ are defined as in (\ref{f1}, \ref{f2}).

Note that the metrics given in equation (\ref{met}) are all locally flat, 
vacuum solutions, allowing possible Dirac delta function type singularities.
The metric is locally flat ,  since we can define new coordinates, 
\begin{equation}      
    d \tau = e^{g_1(x_0+ x_3)} {{(dx_0+ dx_3)} 
   \over {\sqrt{2}}} 
\end{equation} 
\begin{equation}       
 d \overline \tau = e^{g_1(x_0- x_3)} {{(dx_0-dx_3)} 
         \over {\sqrt{2}}} 
\end{equation} 
\begin{equation}      
d \zeta= e^{g_2(x_1+i x_2)} {{ (dx_1+i dx_2)} 
           \over {\sqrt{2}}}
\end{equation} 
\begin{equation}       
d \overline \zeta = e^{g_2(x_1-i x_2)} {{ (dx_1-i dx_2)} 
           \over {\sqrt{2}}} .
\end{equation} 
   Then 
\begin{equation}      
ds^2=d\tau d\overline \tau- d\zeta d\overline \zeta ,
\end{equation} 
a metric which is locally flat.  Only zeros and singularities of 
$g_1$ and $g_2$ can introduce curvature at certain points, giving rise to 
Dirac delta function type singularities.  For particular choices of $g_1$ 
and $g_2$, we see that to perform the transformation from $\zeta$ 
and $\tau$ back to $x_1+ix_0$ and
$x_0+x_3$,  we have to introduce cuts to the $\tau$ and $\zeta$ planes.


For a given metric with of the form given in (6), with a general function 
$f_1$ and $f_2$ the Dirac operator can be derived using standard procedure.
$$  
D=e^{{{f_1(x_0,x_3)} \over {2}}} 
\big[\gamma^0 \partial_0 +\gamma^3 \partial _3 
-{{\gamma^0} \over {4}} {{\partial_0 f_1} }
+{{\gamma^3} \over {4}} {{\partial_3 f_1} } \big]$$
\begin{equation}  
\qquad +e^{{{f_2(x_1,x_2)} \over {2}}}
 \big[\gamma _1 \partial ^1+ \gamma^2 \partial ^2 + {{\gamma^1} \over {4}} 
 {{\partial_1 f_2 } }
 +{{\gamma^2} \over {4}} {{\partial_2 f_2} } \big] .
\end{equation} 
 
 For this general case, the square of the Dirac operator does not give us
the d'Alembertian.  There are extra terms proportional to derivatives
of $f_1$ and $f_2$.  We state this fact by writing
\begin{equation}   
 \Box \not= {Tr\over{4}} DD. 
\end{equation} 
 If we choose $f_1=0$ and $f_2$ so that it can be written as in eq.(5), 
we get 
\begin{equation}   
 {{\gamma^1} \over {4}} {{\partial g_2} \over {\partial x_1}}
 +{{\gamma^2} \over {4}} {{\partial g_2} 
 \over {\partial x_2}}\equiv 0 . 
\end{equation} 
 Then $D'=D$ for $f_1=0$ and eq. (\ref{met}) is satisfied with $D$ as well 
for this special case. To derive this result we used the series expansion 
for $g_2$  and the relation $\gamma^1 \gamma^2 \gamma^1 \gamma^2 =-1$.  
Note that if we take a metric with the Euclidean , instead of the Lorentzian 
signature, this behaviour
can also be generalized to the case where $f_1$ is an arbitrary function, in
the form given as in eq. (4).
With this choice for $f_1$ and $f_2$ the Dirac operator 
is the true square root of the d'Alembertian, similar to its behaviour 
in the free case.  
We will not treat this case in our work, though, and take $f_1=0$.

Our examples  are extensions of the free case, since 
our metrics are all locally flat.  One should recall that solutions with 
only topological defects also have this property. This gives us the 
possibility of finding interesting solutions which generalize
the cosmic string solutions and calculating vacuum fluctuations
in their background.
Below, using semi-classical methods, we will study two cases where   
the scalar and the spinor field is coupled to the metric.
We explicitly study the simplest case which corresponds to a 
'dipole'.  The other cases can be treated along similar lines.

Note that if we do not want to extend  our calculations to spinors, 
we can replace $\gamma^1 \gamma^2$ by $i=\sqrt{-1}$ and $\gamma^0 
\gamma^3$ by unity, where both are multiplied by the unit matrix.  

\section{Scalar field}
A.N.Aliev $^{/3  }$  has shown that at first order perturbation theory, the 
cosmic string gives a  finite contribution to the vacuum energy.  This result 
is in accordance with the expectations, since it is well known that even the 
exact result does not introduce additional infinities. Our result is the 
extension of the Aliev result. The multiple cosmic string solution is a 
special case of our case since we can write 
\begin{equation}   
  \sum_{i=1} ^n \log (x_{1i} ^{2} + x_{2i} ^{2}) = {Tr \over {4}} \left(
   \sum_{i=1} ^{n} (\log ( x_{i1} +\gamma^1 \gamma^2 x_{i2} )+
   \log (x_{i1} - \gamma^1 \gamma^2 x_{i2} )) \right), 
\end{equation} 
where the LHS of the equation corresponds to the multiple-cosmic string
solution $^{/4  }$.   The vacuum fluctuations for the multiple cosmic string
are calculated in references 3 and 5.

 We can show that the similar finite result in first order perturbation 
theory  can be obtained if $f_1, f_2 $ are taken in the form given in 
eq.s(4,5). Then the d`Alembertian operator is of the form given in 
equation(3). 

To illustrate how this mechanism works,   
we will use this formalism to calculate the vacuum fluctuations for one 
special form, which can be interpreted as cosmic strings which result in 
angle defects and excesses.  The same cancellation mechanism works for
the other cases as well.

\bigskip
In our example we take $f_1=0$
\begin{equation} 
g_2 = -\beta \log {(1+\alpha (z-{1\over{z}}))}
\label{g2}
\end{equation} 
When we study only the scalar particle case, we can take $z=x_1+ix_2$.  
If we want to extend our problem also to the spinor case, we replace $z$ 
by $\zeta=x_1+\gamma^1 \gamma^2 x_2$.
and use the $Trace$ operation at appropriate points.
\noindent
We write
\begin{equation}  
 \Box = \partial_0^2 -\partial_3^2-4 e^{\beta \log
(1+\alpha(z-{1\over {z}}))+
\beta  \log (1+\alpha(\overline z -{1\over {\overline z}}))} 
{{\partial} \over {\partial \overline z}} {{\partial } \over 
{ \partial  z}} 
\end{equation} 
In first order perturbation theory we first expand the exponential 
and then the logarithm.
We end up with
\begin{equation}  
 \Box_1 = \partial_0^2-\partial_1^2 -\partial_2^2 - 
\partial_3^2 -\beta \alpha
(2x_1-{{2x_1} \over {x_1^2+x_2^2}})(\partial_1^2 +\partial_2^2). 
\end{equation} 
The first order Greens Function reads
\begin{equation} 
G_F^{(1)}(x-y) = \int dw G_F^{(0)}(x-w) V(w) G_F^{(0)}(w-y) 
\end{equation} 
where
\begin{equation} 
 V=V_1(x_1,x_2)(\partial_1^2+\partial_2^2)   
\end{equation} 
and
\begin{equation}  
 V_1(x_1,x_2)=-\beta \alpha (2x_1 -{2x_1 \over {x_1^2+x_2^2}}). 
\label{pot}
\end{equation} 
If we go to momentum space we have to calculate
\begin{equation}  
 \int dq \int dp {{p_1^2+p_2^2} \over {p^2(p-q)^2}} e^{iqx} V(q) 
\end{equation} 
to obtain the $G_F^{(1)}$ at the coincidence limit .To get $<T_{00}>$, 
the VEV of the energy density, we have to differentiate $G_F^{(1)}$.  
This operation can be shown to result in the integral
\begin{equation}  
 \int dq \int dp {{(p_1^2+p_2^2)^2}\over {p^2(p-q)^2}} e^{iqx} V(q) 
\end{equation} 
where
\begin{equation} 
V(q)=\delta(q_0) \delta (q_3) \int dx_1 dx_2 V_1(x_1,x_2) 
e^{i(q_1 x_1+q_2 x_2)} 
\end{equation} 
$V_1(x_1,x_2)$  was defined in eq. (\ref{pot}).

As Aliev has shown $^{/3}$ this calculation boils down to multiplying 
$V_1(q)$ by $(q_1^2+q_2^2)^2$ and anti Fourier transforming the result.
This gives us
\begin{equation}  
 <T^{00}> = -2A_1{{\alpha \beta x_1}\over {(x_1^2+x_2^2)^3}}
\end{equation} 
for $x_1$ very much smaller than unity, $A_1$ a finite constant, 
$\alpha, \beta$ as in eq.(21).

There are no infinities in this order, since the divergence of
the integral
\begin{equation}  
\int d^4 p {{(p_1^2+p_2^2)}\over { p^2(p-q)^2}} 
\end{equation} 
has been cancelled exactly by the zero of the anti Fourier transforming 
the result. Here we have taken  formulae like
\begin{equation}  
 \int d^2 p e^{ip_1x_1+ip_2x_2} (p_1^2+p_2^2) = 
C {{\epsilon} \over {(x_1^2+x_2^2)^2}} 
\end{equation} 
\begin{equation} 
\int d^2 p e^ {ip_1x_1+ip_2x_2} \log (p_1^2+p_2^2) = 
C' (x_1^2+x_2^2)^{-1}(1+C_1 \epsilon \log (x_1^2+x_2^2)) 
\end{equation} 
as given in $^{/ 6  }$, and used dimensional regularization.  
Here $\epsilon$ is the parameter that goes to zero in the dimensional 
regularization and $C,C',C_1$ are three finite constants.  

Our example, with the choice for $g_2$ given by equation (\ref{g2})
corresponds to three cosmic strings.  Two cosmic strings which give 
rise to defect angles located at $x_1$ and $x_2$ and one cosmic string
which which results in an excess angle, located at the origin. Here $x_1$ and
$x_2$ are given as
\begin{equation}
x_1=-{1\over {2\alpha}} 
-{1\over {2 \alpha}} \sqrt{1+4 \alpha ^2},\qquad x_2=0, 
\end{equation}
\begin{equation} 
x_1=-{1\over {2 \alpha}} + {1\over {2 \alpha}} 
\sqrt {1+ 4 \alpha^2}, x_2=0.\nonumber
\end{equation} 
We have in total one cosmic string and a 
"dipole" made out of cosmic strings.

Although we have explicitly treated one specific example, one can show that
the divergence cancellation occurs in the other cases where equations 
$f_1$ and $f_2$ are satisfied as well. We have to
fourier transform $V_1$ first. This operation may produce $\epsilon$ if
$V_1$ is a monomial in $x^2$. If we have negative powers of $x^2$, we may 
not get an $\epsilon$; however, then we get positive powers of $q^2$.  
Upon anti fourier transforming this expression, we get the necessary 
$\epsilon$ factor multiplying our expressions.  We have exprienced that, 
if we do not take $f_1$ and $f_2$ as given in equations (4) and (5), 
we do not get the necessary power of $\epsilon$ to cancel the one coming 
from the $p$ integration. The metric is not necessarily locally flat then, 
though, and we do not anticipate  his cancellation in the first place.

\section{Spin 1/2 case}

We can calculate the vacuum fluctuation of a spinor in the presence of the 
same configuration. We find that the energy of the spinor particle is of 
the same form as the scalar case.

The vacuum expectation value of energy-momentum tensor for spin 1/2 case 
is given by$^{/7}$
\begin{equation}
< T_{\mu\nu}(s={1\over 2})>=\,{i\over 2}
<\big[{\overline\psi}(x)\gamma_{(\mu}\nabla_{\nu)}\psi(x^\prime)
-(\nabla_{(\mu}{\overline\psi}(x^\prime)\gamma_{(\nu)}\psi(x^\prime)\big]>
\end{equation}
for Dirac spinors. The Feynman propagator $S_F$ is given as a time ordered 
product 
\begin{equation}
S_F(x, x^\prime)=\,i<0 \vert T({\overline\psi}(x^\prime)\psi(x))\vert 0>.
\end{equation} 
For the vacuum expectation value of the energy momentum tensor we have the 
relation
\begin{equation}
< T_{00}(x)>=-Tr[\partial_0 S_F].
\end{equation}
Feynman Green function $S_F(x, x^\prime)$ satisfies the relation
\begin{equation}
i\gamma^\mu(\partial_\mu-\Gamma^\mu)S_F(x,x')=\,{1\over\sqrt{-g}}\delta^4(x-x^\prime) 
\end{equation}
where $\Gamma_\mu$ are spin connection terms for the curved space-time.
In eq.(6) the Dirac operator has the form
\begin{equation}
D=\gamma^0\partial_0+\gamma^3\partial_3+e^{f_2/2}\big[
\gamma^1\partial_1+\gamma^2\partial_2-{1\over 4}(\gamma^1\partial_1-
\gamma^2\partial_2)\big]
\label{dir}
\end{equation}
and if we take $f_2$ as a function of $f_2(x_1-\gamma^1\gamma^2x_2)$ last 
term in equation(\ref{dir}) vanishes and square of the Dirac equation gives 
the d'Alembertian of the spin-0 case:
\begin{equation}
D^2=\partial^2_0-\partial^2_3-e^{(f_2+{\overline f_2})\over2}
(\partial^2_1+\partial^2_2)
\end{equation}
Now spin-1/2 Green function $S_F$ obeys the equation
\begin{equation}
i\big[\gamma^0\partial_0+\gamma^3\partial_3+e^{f_2/2}( 
\gamma^1\partial_1+\gamma^2\partial_2)\big] S_F={1\over e^{-f_2}}\delta^4
(x,x^\prime)
\end{equation}
This expression can be written in the form of
$$
i\gamma^\mu\partial_\mu S_F=\delta^4(x,x^\prime)+i
\big[(1-e^{-f_2})(\gamma^0\partial_0+\gamma^3\partial_3)\nonumber
$$
\begin{equation}
+(1-e^{-f_2/2})(\gamma^1\partial_1+\gamma^2\partial_2)\big] S_F
\end{equation}
which is more convenient for the perturbative approach. 
In the perturbation series, the solution of the equation (38) can be given
as
\begin{equation}
S_F(x,x^\prime)=S_F^{(0)}(x,x^\prime)+S_F^{(0)}V S_F^{(0)}+\cdots
\end{equation}
with  
\begin{equation}
V=i f_2\big(\gamma^0\partial_0+\gamma^3\partial_3+{1\over 2}(
\gamma^1\partial_1+\gamma^2\partial_2)\big).
\end{equation}
$S_F^{(0)}$ is the flat space-time spin 1/2 Green function and has the form
\begin{equation}
S_F^{(0)}(x,x^\prime)=\int{d^4k\over (2\pi)^4}{\gamma k\over k^2}
e^{-ik(x-x^\prime)}.
\end{equation}
 First order correction to the Green function can be evaluated from 
$S_F^{(0)}$
\begin{equation}
S_F^{(1)}=S_F^{(0)}V S_F^{(0)}=\int S_F^{(0)}(x,x^\prime) V(x^{\prime\prime})
S_F^{(0)}(x^{\prime\prime},x^\prime) d^4x^{\prime\prime}.
\end{equation}
and in the coincidence limit $x\rightarrow x^{\prime}$
$$
S_F^{(1)}=\int{d^4kd^4l\over (2\pi)^6}{\gamma^\mu k_\mu}
e^{-ilx} f_2(l)\big[(\gamma^0k_0+\gamma^3 
k_3)+{1\over 2}(\gamma^1(k_1-l_1))$$
\begin{equation}
+\gamma^2(k_2-l_2){\gamma^\nu (k-l)_\nu\over{k^2(k-l)^2}}\delta(l_0)
\delta(l_3)\big]
\end{equation}
and $<T_{00}^{(1)}>$ is
$$
<T_{00}^{(1)}>=-Tr \int{d^4kd^4l\over (2\pi)^6}{\gamma^0 k_0}
{\gamma^\mu k_\mu} e^{-ilx} f_2(l)\big[(\gamma^0k_0+\gamma^3 k_3)$$
\begin{equation}
+{1\over 2}(\gamma^1(k_1-l_2))
+\gamma^2(k_2-l_2){\gamma^\nu (k-l)_\nu\over{k^2(k-l)^2}}\delta(l_0)
\delta(l_3) \big]
\end{equation}
By using the dimensional regularization scheme we get non-zero result for
the vacuum expectation value of the energy-momentum tensor
$$
<T_{00}^{(1)}(x)>=-{4\over{(2\pi)^6}}\big[(-{1\over 2}{\overline f_2}(l)+
f_2(l)){\Gamma({\epsilon\over 2}-1)\over\Gamma(4)}
$$
\begin{equation}
+5f_2(l){\Gamma({\epsilon\over 2}-2)\over\Gamma(6)}\big]\int d^2l(l^2)
^{2-{\epsilon\over 2}}e^{-il}
\label{en1}
\end{equation}
Here we have used the definitions of the functions of $f_2(x)$ and
${\overline f_2}(x)$ 
\begin{equation}
f_2(x)=\beta\log(1+\alpha((x_1-\gamma^1\gamma^2 x_2)-{1\over{(x_1-
\gamma^1\gamma^2 x_2)}})
\end{equation}
and
\begin{equation}
{\overline f_2}(x)=\beta\log(1+\alpha((x_1+\gamma^1\gamma^2 x_2)-
{1\over{(x_1+\gamma^1\gamma^2 x_2)}})
\end{equation}
In equation (\ref{en1}) $f_2(l)$ and ${\overline f_2}(l)$is the Fourier 
transformation of the functions $f_2(x)$ and ${\overline f_2}(x)$ 
respectively. 
Our calculations have been done for the small parameter $\beta$ and $\alpha$.
The expansion of the logarithm to the first order in $\alpha$ gives us 
$f_2(x)$ and ${\overline f_2}(x)$:
\begin{equation}
f_2(x)\approx\beta\log(1+\alpha((x_1-\gamma^1\gamma^2 x_2)-{1\over{(x_1-
\gamma^1\gamma^2 x_2)}})
\end{equation}
and
\begin{equation}
{\overline f_2}(x)\approx\beta\log(1+\alpha((x_1+\gamma^1\gamma^2 x_2)-
{1\over{(x_1+\gamma^1\gamma^2 x_2)}})
\end{equation}
Finally substituting the Fourier transformation of these functions in
(\ref{en1}) we get non-zero result for $<T_{00}^{(1)}>$
\begin{equation}
<T_{00}^{(1)}(x)>=-{\beta\alpha\over{8\pi}}{x_1\over{(x_1^2+x_2^2)^3}}
\end{equation}
This shows the expected behaviour in the stringy space-time.

\section{Conclusion}
Here we have shown how perturbative work on multiple cosmic strings can be 
extended to other locally flat metrics and explicitly calculated the dipole
case for the scalar and the spinor cases.

The same formalism can be applied to other configurations.  
A simple extension will be 
\begin{equation} 
g_2=-\beta \log \left( 1+\alpha (z^2-\frac{1}{ z^2}) \right)\nonumber
\end{equation} 
which corresponds to six cosmic strings, two equally spaced 
on the $x_1$ axis on two sides of the origin,  two more equally spaced on 
two sides of the origin on the $x_2$ axis, and two cosmic strings which result 
in excess angles, located on the origin of the $x_1-x_2$ plane. Then
$<T_{00}>$ will be proportional to 
$$\frac{1}{(x^2+y^2)^{3}}[1-\frac{2x^2}{x^2+y^2}].$$
The same expression will also appear in the $\alpha ^2$ term in the 
expansion of the dipole case.

 By playing with the arbitrary functions one can get more interesting cases,
however, it is not so easy to obtain other than Dirac delta function type 
singularities for the Ricci scalar.
Our metrics are just examples of
locally flat metrics.  This behaviour is also seen in the fact that we can
take the ``square root'' of the d'Alembertian operator in this case. If we 
use the Euclidean metric this is also possible for $f_1\neq 0$ when 
$g_1$ is choosen in the form given by equation 4.
This would allow the study of "non-static" cosmic string-like metrics,
if we can interprete "euclidean time" as time.

At this point one may note that solutions with torsion,
as well as curvature singularities may be obtained $^{/8}$.  It may be
interesting  to generalize our metrics in this direction.

\noindent
{\bf{ Ackowledgement}}
We thank Prof.Dr. A.N. Aliev for giving his results and calculations 
available to us prior to publication and for extensive discussions. 
Discussions with Prof.s A. B\" uy\" ukaksoy, A.H. Bilge, \" O. F. Day\i, 
Y. Nutku are also gratefully acknowledged.  This work is partially supported 
by the the Scientific and Technical Research Council of Turkey and M.H.'s 
work is also  supported by the Academy of Sciences of Turkey.
\vfill\eject
\noindent
{{\bf {REFERENCES}}
\begin {description}
\item {1.} T.W.B. Kibble, Phys. Reports {\bf{67}} (1980) 183, 
M.B. Hindmarsh and
T.W.B. Kibble, Reports  Progress Phys. {\bf{58}} (1995) 477, 
A. Vilenkin and E.P.S. Shellard, {\it{ Cosmic Strings and Other 
Topological Defects}}, Cambridge Univ. Press, Cambridge,  1994;

\item {2.}  T.M. Helliwell and D.A. Konkowski, Phys. Rev. 
{\bf{D34}} (1986) 1908; 
B. Linet, Phys. Rev {\bf{D33}} (1986) 1833, 
Phys. Rev {\bf{D35}} (1987) 536,
A.C.Smith in {\it{ The Formation and Evolution of Cosmic Strings}}, 
Ed. by G.W. Gibbons, S.W.Hawking and T.Vachaspati, Cambridge University 
Press,  Cambridge (1990), p.263;

\item {3.}  A.N.Aliev, Phys.Rev.D {\bf{55}} (1997) 3903;

\item {4.}  P.S. Letelier, Classical and Quantum Gravity 
{\bf{4}} (1987) L75;
single cosmic string solutions can be found at  
J.R. Gott III, Astrophys. J {\bf{288}} (1985) 442, W.A.Hiscock,
Phys.Rev.D {\bf{31}} (1985) 3288, B.Linet, Gen.Rel. Grav. {\bf{17}}
(1985) 1109.

\item {5} A.N.Aliev, M.Horta\c csu and N.\" Ozdemir, 
I.T.U. preprint (1997), submitted to Class. and Quantum Gravity.

\item {6.} I.M.Gelfand and G.E. Shilov, 
{\it{ Generalized Functions}} Vol.1 p.363-364,
Translated by E. Saletan, Academic Press, New York and London, 1964.

\item {7} N.D.Birrell and P.C. Davies, {\it{ Quantum Fields in Curved Space}},
Cambridge University Press, Cambridge, 1982.

 \item {8} K.P.Tod, Class. Quant. Grav. {\bf{11}} (1994) 1331.

\end{description}
\end {document}